\documentclass[8pt]{revtex4}
\pdfoutput=1
\usepackage{amssymb}
\usepackage{latexsym}
\usepackage{epsfig}
\begin{document}                             

\title{ Birth and growth of nonlinear massive gravity and it's transition to nonlinear electrodynamics in a system of Mp-branes  }

\author{Alireza Sepehri $^{1,2}$\footnote{alireza.sepehri@uk.ac.ir} ,  Ahmed Farag Ali$^{3}$\footnote{
ahmed.ali@fsu.edu.eg } }
\address{$^1$ Faculty of Physics, Shahid Bahonar University, P.O. Box 76175, Kerman, Iran.\\$^{2}$
Research Institute for Astronomy and Astrophysics of Maragha
(RIAAM), Maragha, Iran.
\\ $^3$  Department of Physics, Faculty of Science, Benha University, Benha 13518, Egypt.
}

\begin{abstract}
Recently, an interesting mechanism [Phys.Rev.Lett.106:231101,2011]
has been proposed which produces all nonlinear terms in massive
gravity to all orders . In this work, we reproduce these results
in M-theory and consider the process of birth and growth of
nonlinear gravity during the construction of Mp-branes. It has
been shown that Mp brane are built up of p- M1-branes which each
of them are connected to M1-branes of other Mp-brane through a
wormhole. In this model, by increasing the number of dimensions,
the number of nonlinear terms in relevant action of branes
enhances and some theories like lovelock and nonlinear gravity are
raised. By compacting M-branes, graviton fields in nonlinear
gravity
converts to photon fields and thus nonlinear electrodynamics are produced.
\end{abstract}

 \maketitle
\section{Introduction}
A recent interesting idea \cite{w1} has been proposed recently
which suggest that a four-dimensional non-linear theory of massive
gravity, which is a ghost-free in the decoupling limit to all
orders, could  include explicitly all nonlinear terms of an
effective field theory of a massive gravity. There have been
implications with different aspects such as  cosmological
implications  of this idea have been studied as a proxy which
embodied key features of the  non-linear covariant model
\cite{w2}. In the same context of nonlinear massive gravity,
another scenario investigated open Friedmann-Robertson-Walker
(FRW) universes obtained by arbitrary matter source and it has
been derived  three independent branches of solutions. One of them
does not allow for any nontrivial FRW cosmologies and the other
two branches allow for general open FRW universes governed by the
Friedmann equation with the matter source \cite{w3}. Another
approaches have tried to understand the landscape of vacua in
nonlinear massive gravity and considered tunneling between each
pair of adjacent vacua. Several aspects are then studied  such as
the Hawking-Moss (HM) instanton that is located at a local maximum
of the potential, and then investigated the dependence of the
tunneling rate on the parameters of this gravity \cite{w4}.  In
another approach, it has been shown that all homogeneous and
isotropic solutions in nonlinear massive gravity are unstable. It
has been shown that there is at least one ghost that cannot be
removed from the low energy effective theory \cite{w5}. In another
aspect, nonlinear massive gravity has been suggested which
includes F(R) modifications and inherited the benefits of previous
model. This theory  is free of the ghost due to the existence of a
Hamiltonian constraint accompanied by a nontrivial secondary one
\cite{w6}.

Now, a natural question arises on what is the origin of nonlinear
gravity. We try to answer this question in M-theory. In previous
studies, it has been shown that Big Bang may be removed in
M-theory and replaced by k fundamental
strings\cite{w7,w8,w9,w10,w11,ww12,q6,q7}. In this model, k fundamental
strings decay to k pairs of M0-branes. Then, these branes join to
each other and form a system of M3-brane, anti-M3-brane  in
additional to a wormhole. This system is named a BIon. Our
universe is located on one of these M3-branes and interact with
other universe via a wormhole. This wormhole is a channel for
flowing energy from another universe into our universe and hence
occurring an inflation epoch \cite{w7}. In this research, we will
extend these calculations and show that by increasing dimensions
of branes, they can have more interactions with other branes and
more wormholes are formed between them. These wormholes  are
originated from graviton fields in the relevant action of branes
and are the main cause of emergence of nonlinear gravity in four
dimensional universe. If these branes are compacted, some graviton
fields transit to photon fields and nonlinear electrodynamics are
produced.

The outline of the paper is as  follows.  In section \ref{o1}, we
show that by adding dimensions of branes, more graviton fields are
produced which leads to formation of wormholes. These wormholes
are the main cause of generation of nonlinear gravity. In section
\ref{o2}, we will argue that by compacting branes, gravitons
change to photons and nonlinear electrodynamics is produced. The
last section is devoted to summary and conclusion.

 The outline of the
paper is as  follows.  In section \ref{o1}, we will show that by
adding dimensions of branes, more graviton fields are produced
which leads to formation of wormholes. These wormholes are the
main cause of generation of nonlinear gravity. In section
\ref{o2}, we will argue that by compacting branes, gravitons
change to photons and nonlinear electrodynamics is produced. The
last section is devoted to summary and conclusion.

\section{ Emergence of nonlinear gravity in an Mp-brane
}\label{o1} Previously, it has been shown that all branes can be
constructed from M0-branes \cite{w7}. These branes have no gauge
fields and only scalars interact with them. By adding a dimension
to the brane, the first gauge field is appeared and attached to
M1. This field plays the role of graviton and produces linear
gravity. Increasing dimensions, more graviton fields stick to
brane and cause to generation of nonlinear gravity. This gravity
leads to emergence of a large number of wormholes which connect a
brane to other branes. Branes and wormholes in this system form a
complicated system of BIons. To begin, we introduce the Lagrangian
of nonlinear gravity \cite{w1,w2}:

\begin{equation}
L=\frac{M_{P}^{2}}{2}\sqrt{-g}\big(R-\frac{m_{g}^{2}}{4}U(g,H)\big)\label{a1}
\end{equation}
where $U$ is the gravitational potential, $m_{g}$ is the graviton
mass, $H_{\mu\nu}$ is constructed in terms of the metric
$g_{\mu\nu}$ and the four Stuckelberg fields $\Phi^{a}$ by
$H_{\mu\nu}=
g_{\mu\nu}-\eta_{ab}\partial_{\mu}\Phi^{a}\partial_{\nu}\Phi^{b}$.
The most generic potential, U, has the following form \cite{w2}
\begin{equation}
U=-4(U_{2}+\alpha_{3} U_{3}+\alpha_{4}U_{4}), \label{a2}
\end{equation}
where $\alpha_{3,4}$ are two free parameters and
\begin{eqnarray}
&& U_{2}=-\frac{1}{2}\varepsilon_{\mu\alpha\rho\sigma}\varepsilon^{\nu\beta\rho\sigma}\kappa^{\mu}_{\nu}\kappa^{\alpha}_{\beta}=[\kappa]^{2}-[\kappa^{2}],  \nonumber\\
&&U_{3}=-\varepsilon_{\mu\alpha\gamma\rho}\varepsilon^{\nu\beta\delta\rho}\kappa^{\mu}_{\nu}\kappa^{\alpha}_{\beta}\kappa^{\gamma}_{\delta}=[\kappa]^{3}-3[\kappa^{2}][\kappa]+2[\kappa^{3}],
\nonumber\\
&&U_{4}=-\varepsilon_{\mu\alpha\gamma\rho}\varepsilon^{\nu\beta\delta\sigma}\kappa^{\mu}_{\nu}\kappa^{\alpha}_{\beta}\kappa^{\gamma}_{\delta}\kappa^{\rho}_{\sigma}=[\kappa]^{4}-6[\kappa^{2}][\kappa]^{2}+8[\kappa^{3}]
[\kappa]+3[\kappa^{2}]^{2}-6[\kappa^{4}],     \label{a3}
\end{eqnarray}
where
\begin{eqnarray}
&&\kappa^{\mu}_{\nu}=\delta^{\mu}_{\nu}-\sqrt{\delta^{\mu}_{\nu}-H^{\mu}_{\nu}}\nonumber\\
&&H_{\mu\nu}=h_{\mu\nu}+2\Pi_{\mu\nu}-\eta^{\alpha\beta}\Pi_{\mu\alpha}\Pi_{\beta\nu}\nonumber\\
&&\Phi^{a}=x^{a}-\eta^{a\mu}\partial_{\mu}\pi \nonumber\\
&&\Pi_{\mu\nu}=\partial_{\mu}\partial_{\nu}\pi \label{aa3}
\end{eqnarray}

 We can show this
Lagrangian can be extracted from relevant actions of M-branes in a
BIonic system. In fact, only one graviton field couples to M1 and
produces a linear gravity. Also, this field generates a wormhole
which connect M1-brane to other branes. By joining two M1-branes
to each other, one M2-brane is constructed which two gravitons
attach to it. These gravitons leads to first signature of
nonlinear massive gravity and appearance of two wormholes. By
joining M1-branes, dimension of brane increases and more gravitons
are produced and cause to the emergence of a lot of wormholes and
nonlinear gravity. The action of M1 can be written as
\cite{w7,q6,q7}:

\begin{eqnarray}
&& S = - T_{M1}\int  d^{2}\sigma ~ STr \Bigg(-det(P^{mnl}([ E_{mnl}
+E_{mij}(Q^{-1}-\delta)^{ijk}E_{kln}]+\nonumber
\\&& \lambda F_{mnl}))det(Q)\Bigg)^{1/2}~~ \label{a4}
\end{eqnarray}

where

\begin{eqnarray}
  && E_{mnl}^{\alpha,\beta,\gamma} = G_{mnl}^{\alpha,\beta,\gamma} + B_{mnl}^{\alpha,\beta,\gamma}, \qquad Q=Q^{i}_{ii} \qquad Q^{i}_{j,k} = \delta^{i}_{j,k}
    + i\lambda[X^{j}_{\alpha}T^{\alpha},X^{k}_{\beta}T^{\beta},X^{k'}_{\gamma}T^{\gamma}]E_{k'jl}^{\alpha,\beta,\gamma}\nonumber\\&&F_{abc}=\partial_{a} A_{bc}-\partial_{b} A_{ca}+\partial_{c}
A_{ab} \label{a5}
\end{eqnarray}

Here $X^{M}=X^{M}_{\alpha}T^{\alpha}$, $A_{ab}$ is 2-form gauge
field,

\begin{eqnarray}
 &&[T^{\alpha}, T^{\beta}, T^{\gamma}]= f^{\alpha \beta \gamma}_{\eta}T^{\eta} \nonumber \\&& [X^{M},X^{N},X^{L}]=[X^{M}_{\alpha}T^{\alpha},X^{N}_{\beta}T^{\beta},X^{L}_{\gamma}T^{\gamma}]
\label{a6}
\end{eqnarray}

 $\lambda=2\pi l_{s}^{2}$,
$G_{mnl}=g_{mn}\delta^{n'}_{n,l}+\partial_{m}X^{i}\partial_{n'}X^{i}\delta^{n'}_{n,l}$
and $X^{i}$ are scalar fields of mass
 dimension. Here $a,b=0,1,...,p$
are the world-volume indices of the Mp-branes, $i,j,k = p+1,...,9$
are indices of the transverse space, and m,n are the
ten-dimensional spacetime indices. Also,
$T_{Mp}=\frac{1}{g_{s}(2\pi)^{p}l_{s}^{p+1}}$ is the tension of
Mp-brane, $l_{s}$ is the string length and $g_{s}$ is the string
coupling. This action has been proposed for the first time in \cite{w7,q6,q7}for  interactions with three li-algebra in M-theory and is very different of usual action \cite{w12,w13,w14,w15,w16} in string theory which obey of two li-algebra.

Scalars and two form gauge fields should satisfy following
relations \cite{w7,q6,q7,w17,w18,w19,w20}:

\begin{eqnarray}
&&\langle[X^{a},X^{b},X^{i}],[X^{a},X^{b},X^{i}]\rangle=
\frac{1}{2}\varepsilon^{abc}\varepsilon^{abd}(\partial_{a}X^{i}_{\alpha})(\partial_{a}X^{i}_{\beta})\langle(T^{\alpha},T^{\beta}\rangle
=
 \frac{1}{2}\langle \partial_{a}X^{i},\partial_{a}X^{i}\rangle \nonumber \\
&&\nonumber \\
&&\langle[X^{a},X^{b},X^{c}],[X^{a},X^{b},X^{c}]\rangle=
(F^{abc}_{\alpha\beta\gamma})(F^{abc}_{\alpha\beta\eta})\langle[T^{\alpha},T^{\beta},T^{\gamma}],[T^{\alpha},T^{\beta},T^{\eta}]\rangle)=\nonumber \\
&&
(F^{abc}_{\alpha\beta\gamma})(F^{abc}_{\alpha\beta\eta})f^{\alpha
\beta \gamma}_{\sigma}h^{\sigma \kappa}f^{\alpha \beta
\eta}_{\kappa} \langle T^{\gamma},T^{\eta}\rangle=
(F^{abc}_{\alpha\beta\gamma})(F^{abc}_{\alpha\beta\eta})\delta^{\kappa
\sigma} \langle T^{\gamma},T^{\eta}\rangle= \langle
F^{abc},F^{abc}\rangle \nonumber \\
&&\nonumber \\
&& i,j=p+1,..,10\quad a,b=0,1,...p\quad m,n=0,..,10~~ \label{a7}
\end{eqnarray}

In this model, two form gauge field plays the role of tensor state
of graviton  and we can replace it by graviton and obtain
following results:

\begin{eqnarray}
&&A^{ab}=g^{ab}=h^{ab}+\eta^{ab} \quad \text{and} \quad
a,b,c=\mu,\nu,\lambda \Rightarrow \nonumber\\&&
F_{abc}=\partial_{a} A_{bc}-\partial_{b} A_{ca}+\partial_{c}
A_{ab}=2(\partial_{\mu}g_{\nu\lambda}+\partial_{\nu}g_{\mu\lambda}-\partial_{\lambda}g_{\mu\nu})=2\Gamma_{\mu\nu\lambda}\nonumber\\&&\nonumber\\&&\langle
F^{\rho}\smallskip_{\sigma\lambda},F^{\lambda}\smallskip_{\mu\nu}\rangle=\langle[X^{\rho},X_{\sigma},X_{\lambda}],[X^{\lambda},X_{\mu},X_{\nu}]\rangle=\nonumber\\&&
[X_{\nu},[X^{\rho},X_{\sigma},X_{\mu}]]-[X_{\mu},[X^{\rho},X_{\sigma},X_{\nu}]]+\nonumber\\&&[X^{\rho},X_{\lambda},X_{\nu}][X^{\lambda},X_{\sigma},X_{\mu}]
-[X^{\rho},X_{\lambda},X_{\mu}][X^{\lambda},X_{\sigma},X_{\nu}]=\nonumber\\&&\partial_{\nu}\Gamma^{\rho}_{\sigma\mu}-\partial_{\mu}\Gamma^{\rho}_{\sigma\nu}+\Gamma^{\rho}_{\lambda\nu}\Gamma^{\lambda}_{\sigma\mu}-\Gamma^{\rho}_{\lambda\mu}\Gamma^{\lambda}_{\sigma\nu}
=R^{\rho}_{\sigma\mu\nu}\label{a8}
\end{eqnarray}

 These equations show that  curvature tensor easily can be obtained from the noncommutative relations  in M-theory. In fact, by adding dimensions to branes
 and growing gauge fields on them, the first signature of gravity can be detected. Previously in (Physics Letters B 748 (2015) 328335, Eur.Phys.J. C76 (2016)
 no.5, 231.), we have proposed a new model in M-theory which allows for constructing all Mp-branes from M1 and M0-branes. In this published model, we have shown that  for M0-branes, there is no guage field and only scalars have attached to it, however by joining M0-branes and formation of M1-branes, gauge fields emerges on it. These M1-branes are linked to anti-M1-branes and form a new system name BIon. For this system metric can be constructed from metrics of two M1-BIon as follows:
 
 $Metric of BIon=(\textbf{Metric M1})_{1}\times (\textbf{Metric M1})_{2}-(\textbf{Metric M1})_{2}\times (\textbf{Metric M1})_{1}$
 
 Thus metric of BIon can be antisymmetric. On the other hand, tensor mode of graviton has a direct relation with metric of BIon and can be anti-symmetric. Also, we have shown that M-theory is a more complete version of string theory. Usual string theory is a ten dimensional theory and includes several versions like closed string theory and open string theory which each of them considers the universe via itself point of view. However M-theory is an eleven dimensional space-time them theory which is more complete than string theory and has many differences with it. In this theory, two form gauge fields have the main role in phenomenological events between branes. In this model, we have argued that scalars in M-theory which move between branes play the role of scalar mode of graviton and are similar to closed strings that play the role of graviton. Also, two form gauge fields play the role of tensor mode of graviton.
  Using equation (\ref{a4}), we can obtain following relations:

\begin{eqnarray}
&&
\det(Z)=\delta^{a_{1},a_{2}...a_{n}}_{b_{1}b_{2}....b_{n}}Z^{b_{1}}_{a_{1}}...Z^{b_{n}}_{a_{n}}
\quad a,b,c=\mu,\nu,\lambda \nonumber\\\nonumber\\
&&Z^{b_{n}}_{a_{n}}=\delta^{b_{n}}_{a_{n}} P^{mnl}([ E_{mnl} +E_{mij}(Q^{-1}-\delta)^{ijk}E_{kln}]+
\lambda F_{mnl})\nonumber\\&&\nonumber\\ &&\det(Z)=\det(P^{mnl}[
E_{mnl}
+E_{mij}(Q^{-1}-\delta)^{ijk}E_{kln}])+\lambda^{2}\det(F)\label{E1}\end{eqnarray}

This equation helps us to calculate the relevant terms of
determinant in action (\ref{a4}) separately. By substituting equations (\ref{a7}) and( \ref{a8}) into above
determinants we obtain

\begin{eqnarray}
\det(F)=\delta_{\rho\sigma}^{\mu\nu}\langle
F^{\rho\sigma}\smallskip_{\lambda},F^{\lambda}\smallskip_{\mu\nu}\rangle
=\delta_{\rho\sigma}^{\mu\nu}R^{\rho\sigma}_{\mu\nu}\label{E2}\end{eqnarray}

\begin{eqnarray}
&& \det(P^{mnl}[ E_{mnl}
+E_{mij}(Q^{-1}-\delta)^{ijk}E_{kln}])=\nonumber\\&&\delta_{\rho\sigma}^{\mu\nu}
[(g^{\mu}_{\rho}g^{\nu}_{\sigma}+ g^{\nu}_{\sigma}\langle
\partial^{\mu}X^{i},\partial_{\rho}X^{j}\rangle+..)+\nonumber\\&&\frac{(g^{\mu}_{\rho}g^{\nu}_{\sigma}+ g^{\nu}_{\sigma}\langle
\partial^{\mu}X^{i},\partial_{\rho}X^{j}\rangle+..)}{[(\lambda)^{2}\det([X^{j}_{\alpha}T^{\alpha},X^{k}_{\beta}T^{\beta},X^{k'}_{\gamma}T^{\gamma}])]}]=
\nonumber\\&&\delta_{\rho\sigma}^{\mu\nu}\kappa_{\mu}^{\rho}\kappa_{\nu}^{\sigma}
(1+\frac{1}{[(\lambda)^{2}\det([X^{j}_{\alpha}T^{\alpha},X^{k}_{\beta}T^{\beta},X^{k'}_{\gamma}T^{\gamma}])]})=
\nonumber\\&&\delta_{\rho\sigma}^{\mu\nu}\kappa_{\mu}^{\rho}\kappa_{\nu}^{\sigma}(1+\frac{1}{m_{g}^{2}})\label{E3}
\end{eqnarray}
where
$m_{g}^{2}=[(\lambda)^{2}\det([X^{j}_{\alpha}T^{\alpha},X^{k}_{\beta}T^{\beta},X^{k'}_{\gamma}T^{\gamma}])]
$ is the square of graviton mass. As can be seen from this
definition, the graviton mass depends on the scalars which
interact with branes. This definition helps us to calculate
another term of determinant:

\begin{eqnarray}
 &&\det(Q)\sim
(i)^{2}(\lambda)^{2}\det([X^{j}_{\alpha}T^{\alpha},X^{k}_{\beta}T^{\beta},X^{k'}_{\gamma}T^{\gamma}])\det(E)\sim
\nonumber\\&&
-[(\lambda)^{2}\det([X^{j}_{\alpha}T^{\alpha},X^{k}_{\beta}T^{\beta},X^{k'}_{\gamma}T^{\gamma}])]\det(g)=m_{g}^{2}\det(g)
\label{a66}
\end{eqnarray}

By inserting equations (\ref{E2}) ,(\ref{E3}), (\ref{a66}) into the action
(\ref{a4}) and doing some approximations, we can obtain:

\begin{eqnarray}
&& S_{M1} =  -T_{M1} \int d^{2}\sigma
[\sqrt{-g}\Big(\delta^{\rho\sigma}_{\mu\nu}\kappa^{\mu}_{\rho}\kappa^{\nu}_{\sigma}+
m_{g}^{2}\delta^{\rho\sigma}_{\mu\nu}(R_{\rho\sigma}^{\mu\nu}+\kappa^{\mu}_{\rho}\kappa^{\nu}_{\sigma})\Big)]=\nonumber\\&&
-T_{M1} \int d^{2}\sigma [\sqrt{-g}\Big(R+
m_{g}^{2}\delta^{\rho\sigma}_{\mu\nu}R_{\rho\sigma}^{\mu\nu}+m_{g}^{2}R\Big)]
\label{a9}
\end{eqnarray}

where we have used of this fact that
($\delta^{\rho\sigma}_{\mu\nu}\kappa^{\mu}_{\rho}\kappa^{\nu}_{\sigma}=R$)\cite{w21}.
This equation gives the leading order of terms in massive gravity.
Also, it is clear that the first term of Lovelock gravity appears
in this action. To achieve to higher order of terms in massive and
Lovelock gravity, we should add dimension to branes.  For this,
using equation (\ref{a7}), we apply following replacement in
action of branes.

\begin{eqnarray}
&&i,j=a,b\Rightarrow
\langle[X^{i},X^{j},X^{k}],[X^{i},X^{j},X^{k}]\rangle
\Rightarrow \langle[X^{a},X^{b},X^{i}],[X^{a},X^{b},X^{i}]\rangle= \frac{1}{2}\langle \partial_{a}X^{i},\partial_{a}X^{i}\rangle \nonumber \\
&&\nonumber \\
&&i,j,k=a,b,c\Rightarrow
\langle[X^{i},X^{j},X^{k}],[X^{i},X^{j},X^{k}]\rangle \Rightarrow
\langle[X^{a},X^{b},X^{c}],[X^{a},X^{b},X^{c}]\rangle=  \langle
F^{abc},F^{abc}\rangle \nonumber \\
&&\nonumber \\
&& Q^{i}_{j,k} = \delta^{i}_{j,k}
    + i\lambda[X^{j}_{\alpha}T^{\alpha},X^{k}_{\beta}T^{\beta},X^{k'}_{\gamma}T^{\gamma}]E_{k'jl}^{\alpha,\beta,\gamma}\Rightarrow \nonumber \\
&& Q^{i}_{j,k} = \delta^{i}_{j,k}
    + i\Big(\lambda[X^{j}_{\alpha}T^{\alpha},X^{k}_{\beta}T^{\beta},X^{k'}_{\gamma}T^{\gamma}]+
    \langle[X^{a},X^{b},X^{i}],[X^{a},X^{b},X^{i}]\rangle+\langle[X^{a},X^{b},X^{c}],[X^{a},X^{b},X^{c}]\rangle\Big)E_{k'jl}^{\alpha,\beta,\gamma}=\nonumber \\
&&\delta^{i}_{j,k}
    + i\Big(\lambda[X^{j}_{\alpha}T^{\alpha},X^{k}_{\beta}T^{\beta},X^{k'}_{\gamma}T^{\gamma}]+
    \frac{1}{2}\langle \partial_{a}X^{i},\partial_{a}X^{i}\rangle+\langle
F^{abc},F^{abc}\rangle\Big)E_{k'jl}^{\alpha,\beta,\gamma} \nonumber \\
&&\nonumber \\
&& T_{M1}\int d^{2}\sigma \Rightarrow T_{Mp}\int d^{p}\sigma
\label{a10}
\end{eqnarray}

Substituting the relations in equation (\ref{a10}) into action
(\ref{a4}) we have:

\begin{eqnarray}
&& S=-T_{Mp} \int d^{p}\sigma \sqrt{-det ( O+2\pi
l_{s}^{2}G(F))}\nonumber\\&&
G=\sum_{n=0,..p}\frac{1}{n!}(-\frac{F}{\beta^{2}})^{n}
\nonumber\\&&\nonumber\\&&O=\frac{1}{p}\sum_{n}(p-n)!\frac{Y^{n}}{n!}
\nonumber\\&&\nonumber\\&& F=\langle F^{abc},F^{abc}\rangle\quad
Y=\langle
\partial_{a}X^{i},\partial_{a}X^{i}\rangle \quad \beta=\frac{1}{2\pi
l_{s}^{2}} \label{a11}
\end{eqnarray}

where we have used of the form of nonlinear field (G) in
\cite{ww12,ww13,ww14}. At this stage, we show that this action can
be written in terms of relevant actions of p M1's. Also, we argue
that by adding each dimension, the probability for producing one
wormhole between brane and other branes are enhanced. For
simplicity, we choose
 $X^{1}=\sigma$ and $X^{4}=z$, where z is the transverse direction
between branes.  Using the  above action, the  Lagrangian
for Mp-brane can be written as

\begin{eqnarray}
&&\L=- T_{Mp} \int d\sigma \sigma^{p}\sqrt{(1+ z'^{2})^{p}+(2\pi
l_{s}^{2})^{2}G(F)}\label{a12}
\end{eqnarray}

where ($'$) denotes the derivative respect to $\sigma$. To obtain
the Hamiltonian, we need to derive the canonical momentum density
for graviton. For simplicity, we  will use of the method in
\cite{ww20} and \cite{ww21} and assume that $F_{001}\neq 0$ and
other components of F are zero. We get:

\begin{eqnarray}
&&\Pi=\frac{\delta \L}{\delta
\partial_{t}A^{01}}=\frac{\sum_{n=0}^{p}\frac{n}{n!}(-\frac{F}{\beta^{2}})^{n-1}F_{001}}{\sqrt{(1+
z'^{2})^{p}+(2\pi l_{s}^{2})^{2}G(F)}} \label{a13}.
\end{eqnarray}

Thus the Hamiltonian can be written as:

\begin{eqnarray}
&&H= T_{Mp}\int d\sigma
\sigma^{p}\Pi\partial_{t}A^{01}-\L=4\pi\int d\sigma [
\sigma^{p}\Pi(\partial_{t}A^{01}-\partial_{\sigma}A^{00})-\partial_{\sigma}(\sigma^{2}\Pi)A_{00}]-\L
\label{a14}
\end{eqnarray}

In the second step integrated by parts, we have the term
proportional to $\partial_{\sigma}A^{01}$. Using the constraint
($\partial_{\sigma}(\sigma^{p}\Pi)=0$), we get \cite{ww20}:

\begin{eqnarray}
&& \Pi=\frac{k}{4\pi \sigma^{p}} \label{a15}
\end{eqnarray}

where $k$ is a constant. Substituting $\Pi$ from above equation into
equation (\ref{a14}) yields the following Hamiltonian:

\begin{eqnarray}
&&H_{1}= T_{Mp}\int d\sigma \sigma^{p} \sqrt{(1+z'^{2})^{p}+(2\pi
l_{s}^{2})^{2}\sum_{n=0}^{p}\frac{n}{n!}(-\frac{F}{\beta^{2}})^{n}}F_{1}\nonumber\\&&F_{1}=\sqrt{1+\frac{k^{2}}{\sigma^{2p}}}
\label{a16}
\end{eqnarray}

To derive the explicit form of wormhole between branes, we need
a Hamiltonian which be only in terms of separation distance
between branes. For this reason, we redefine Lagrangian as:

\begin{eqnarray}
&&\L=- T_{Mp} \int d\sigma \sigma^{p}\sqrt{(1+z'^{2})^{p}+(2\pi
l_{s}^{2})^{2}\sum_{n=0}^{p}\frac{n}{n!}(-\frac{F}{\beta^{2}})^{n}}F_{1}\label{a17}
\end{eqnarray}

With this new form of Lagrangian, we repeat our previous
calculations. We can write :

\begin{eqnarray}
&&\Pi=\frac{\delta \L}{\delta
\partial_{t}A^{01}}=\frac{\sum_{n}^{p}\frac{n(n-1)}{n!}(-\frac{F}{\beta^{2}})^{n-1}F_{001}}{\sqrt{(1+z'^{2})^{p}+(2\pi
l_{s}^{2})^{2}\sum_{n=0}\frac{n}{n!}(-\frac{F}{\beta^{2}})^{n}}}
\label{a18}
\end{eqnarray}

Therefore the new Hamiltonian can be built as:

\begin{eqnarray}
&&H_{2}= T_{Mp}\int d\sigma
\sigma^{p}F_{1}\Pi\partial_{t}A^{01}-\L=\int d\sigma [
\sigma^{p}F_{1}\Pi(\partial_{t}A^{01}-\partial_{\sigma}A^{00})-\partial_{\sigma}(F_{1}\sigma^{p}\Pi)A_{00}]-\L
\label{a19}
\end{eqnarray}

where like previous stage, we have in the second step integrated
by parts the term proportional to $\partial_{\sigma}A^{01}$ like
the method in \cite{ww20}. Imposing the constraint
($\partial_{\sigma}(\sigma^{p}\Pi)=0$), we get:

\begin{eqnarray}
&& \Pi=\frac{k}{4F_{1}\pi \sigma^{p}} \label{a20}
\end{eqnarray}

By substituting equation (\ref{a20}) into
equation (\ref{a19}) we obtain the following Hamiltonian:

\begin{eqnarray}
&&H_{2}= T_{Mp}\int d\sigma \sigma^{p}\sqrt{(1+z'^{2})^{p}+(2\pi
l_{s}^{2})^{2}\sum_{n}^{p}\frac{n(n-1)}{n!}(-\frac{F}{\beta^{2}})^{n}}F_{2}\nonumber\\&&F_{2}=F_{1}\sqrt{1+\frac{k^{2}}{F_{1}^{2}\sigma^{2p}}}
\label{a21}
\end{eqnarray}

and by repeating these calculations for $p$ times, we obtain:

\begin{eqnarray}
&&H_{p}= T_{Mp}\int d\sigma \sigma^{p} \sqrt{(1+z'^{2})^{p}+(2\pi
l_{s}^{2})^{2}\sum_{n=0}^{p}\frac{n(n-1)....(n-n)}{n!}(-\frac{F}{\beta^{2}})^{n}}F_{p}\nonumber\\&&
=T_{Mp}\int d\sigma \sigma^{p}
(\sqrt{1+z'^{2}})^{p}F_{tot}\nonumber\\&&
F_{tot}=\sqrt{1+\frac{k^{2}}{F_{p-1}^{2}\sigma^{2p}}}\sqrt{1+\frac{k^{2}}{F_{p-2}^{2}\sigma^{2p}}}...\sqrt{1+\frac{k^{2}}{F_{1}^{2}\sigma^{2p}}}\sqrt{1+\frac{k^{2}}{\sigma^{2p}}}\nonumber\\&&F_{n}=F_{n-1}\sqrt{1+\frac{k^{2}}{F_{n-1}^{2}\sigma^{2p}}}
\label{a22}
\end{eqnarray}

In the case of $\frac{k}{\sigma^{p}}\ll1$, we can reobtain the
Hamiltonian of Mp-brane by multiplying the Hamiltonians of M1's:

\begin{eqnarray}
&&H_{p}= T_{Mp}\int d\sigma \sigma^{p}
(\sqrt{1+z'^{2}})^{p}F_{tot}\nonumber\\&& k=k'^{p}\rightarrow
F_{tot}=\sqrt{1+\frac{k^{2}}{F_{p-1}^{2}\sigma^{2p}}}\sqrt{1+\frac{k^{2}}{F_{p-2}^{2}\sigma^{2p}}}...
\sqrt{1+\frac{k^{2}}{F_{1}^{2}\sigma^{2p}}}\sqrt{1+\frac{k^{2}}{\sigma^{2p}}}\simeq
\nonumber\\&&
\sigma^{p}\sqrt{1+\frac{k'^{2p}}{\sigma^{2p}}+p\frac{k'^{2p-1}}{\sigma^{2p-1}}+..}=\sqrt{(1+\frac{k'^{2}}{\sigma^{2}})^{p}}\Rightarrow
\nonumber\\&& H_{p}= T_{Mp}\int d\sigma \sigma^{2p}
(\sqrt{1+z'^{2}})^{p}(\sqrt{1+\frac{k'^{2}}{\sigma^{2}}})^{p}=T_{Mp}\int
d^{p}\sigma \sigma^{p}
(\sqrt{1+z'^{2}})^{p}(\sqrt{1+\frac{k'^{2}}{\sigma^{2}}})^{p}\Rightarrow
\nonumber\\&& H_{p}=(T_{M1}\int d\sigma \sigma
\sqrt{1+z'^{2}}\sqrt{1+\frac{k'^{2}}{\sigma^{2}}})^{p} =H_{1}^{p}
\nonumber\\&& \nonumber\\&& \nonumber\\&& H_{1}=T_{M1}\int d\sigma
\sigma
\sqrt{1+z'^{2}}\sqrt{1+\frac{k'^{2}}{\sigma^{2}}}\label{a23}
\end{eqnarray}

where we have used of this assumption that $T_{Mp}=(T_{M1})^{p}$.
This equation says that each Mp-brane can be built of p M1-brane.
Also, we can show that each M1-brane produces a wormhole. To this
end, using the above Hamiltonian, we obtain the following equation
of motion  $z$ for any M1:

\begin{eqnarray}
&&-z_{M1}'=(\frac{V_{1}(\sigma)^{2}}{V_{1}(\sigma_{0})^{2}}-1)^{-1/2}
\nonumber\\&& V_{1}=\sigma F_{1}=\sigma
\sqrt{1+\frac{k'^{2}}{\sigma^{2}}}\label{a24}
\end{eqnarray}

The solution of this equation is:

\begin{eqnarray}
&&z_{M1}= \int_{\sigma}^{\infty}
d\sigma'(\frac{V_{1}(\sigma')^{2}}{V_{tot}(\sigma_{0})^{2}}-1)^{-1/2}
\label{a25}
\end{eqnarray}

Thus, the separation distance between two branes can be given by:

\begin{eqnarray}
&&\Delta_{M1}=2z_{M1}= 2\int_{\sigma}^{\infty}
d\sigma'(\frac{V_{1}(\sigma')^{2}}{V_{1}(\sigma_{0})^{2}}-1)^{-1/2}
\label{a26}
\end{eqnarray}

 where $\sigma_{0}$ is the throat of wormhole between two
 M1-branes of two different branes. In fact, these results show
 that by joining p M1-branes, we can construct one Mp-brane. Then
 this brane may interact with other branes via p wormholes.

At this stage, we can obtain the relevant action for Mp-branes by
multiplying the action of p M1-branes. However, we should care in
antisymmetric form $\delta$. We can write:

\begin{eqnarray}
&&S_{Mp} = -T_{Mp} \int dt \L_{Mp}  \nonumber\\&& \L_{Mp}=\det(M)
\quad L_{M1,i}=L^{b_{i}}_{a_{i}}=\det(M_{i})\sim M_{i}
\nonumber\\&&\det(M)=\delta^{a_{1},a_{2}...a_{n}}_{b_{1}b_{2}....b_{n}}M^{b_{1}}_{a_{1}}...M^{b_{p}}_{a_{p}}\Rightarrow
\nonumber\\&&
\L_{Mp}=\det(M)=\delta^{a_{1},a_{2}...a_{n}}_{b_{1}b_{2}....b_{n}}L^{b_{1}}_{a_{1}}...L^{b_{p}}_{a_{p}}
\nonumber\\&&\delta^{a_{1},a_{2}...a_{n}}_{b_{1}b_{2}....b_{n}}\delta^{\rho_{1}\sigma_{1}}_{\mu_{1}\nu_{1}}...\delta^{\rho_{p}\sigma_{p}}_{\mu_{p}\nu_{p}}
=\delta^{\rho_{1}\sigma_{1}...\rho_{p}\sigma_{p}}_{\mu_{1}\nu_{1}...\mu_{p}\nu_{p}}\nonumber\\&&
\sqrt{-g}=\sqrt{-\det(g)}=\sqrt{-\det(g_{1}g_{2}...g_{p})}=\sqrt{-\det(g_{1})\det(g_{2})...\det(g_{p})}\nonumber\\&&\nonumber\\&&\nonumber\\&&
S_{Mp} = -(T_{M1})^{p} \int dt
\smallskip
\delta^{a_{1},a_{2}...a_{n}}_{b_{1}b_{2}....b_{n}}L^{b_{1}}_{a_{1}}...L^{b_{p}}_{a_{p}}=\nonumber\\&&
-(T_{M1})^{p} \int dt \int d^{p}\sigma
\delta^{a_{1},a_{2}...a_{n}}_{b_{1}b_{2}....b_{n}}[\sqrt{-g_{1}}\Big(\delta^{\rho_{1}\sigma_{1}}_{\mu_{1}\nu_{1}}\kappa^{\mu_{1}}_{\rho_{1}}\kappa^{\nu_{1}}_{\sigma_{1}}+
m_{g}^{2}\delta^{\rho_{1}\sigma_{1}}_{\mu_{1}\nu_{1}}(R_{\rho_{1}\sigma_{1}}^{\mu_{1}\nu_{1}}+\kappa^{\mu_{1}}_{\rho_{1}}\kappa^{\nu_{1}}_{\sigma_{1}})\Big)]^{b_{1}}_{a_{1}}\times
\nonumber\\&&
[\sqrt{-g_{2}}\Big(\delta^{\rho_{2}\sigma_{2}}_{\mu_{2}\nu_{2}}\kappa^{\mu_{2}}_{\rho_{2}}\kappa^{\nu_{2}}_{\sigma_{2}}+
m_{g}^{2}\delta^{\rho_{2}\sigma_{2}}_{\mu_{2}\nu_{2}}(R_{\rho_{2}\sigma_{2}}^{\mu_{2}\nu_{2}}+\kappa^{\mu_{2}}_{\rho_{2}}\kappa^{\nu_{2}}_{\sigma_{2}})\Big)]^{b_{2}}_{a_{2}}\times
...\times\nonumber\\&&[\sqrt{-g_{p}}\Big(\delta^{\rho_{p}\sigma_{p}}_{\mu_{p}\nu_{p}}\kappa^{\mu_{p}}_{\rho_{p}}\kappa^{\nu_{p}}_{\sigma_{p}}+
m_{g}^{2}\delta^{\rho_{p}\sigma_{p}}_{\mu_{p}\nu_{p}}(R_{\rho_{p}\sigma_{p}}^{\mu_{p}\nu_{p}}+\kappa^{\mu_{p}}_{\rho_{p}}\kappa^{\nu_{p}}_{\sigma_{p}})\Big)]^{b_{p}}_{a_{p}}=\nonumber\\&&-(T_{Mp})
\int dt \int d^{p}\sigma
[\sqrt{-g}\Big(\sum_{n=1}^{p}\delta^{\rho_{1}\sigma_{1}...\rho_{n}\sigma_{n}}_{\mu_{1}\nu_{1}...\mu_{n}\nu_{n}}\kappa^{\mu_{1}}_{\rho_{1}}\kappa^{\nu_{1}}_{\sigma_{1}}
..\kappa^{\mu_{n}}_{\rho_{n}}\kappa^{\nu_{n}}_{\sigma_{n}}+\nonumber\\&&
\sum_{n=1}^{p}m_{g}^{2n}\delta^{\rho_{1}\sigma_{1}...\rho_{n}\sigma_{n}}_{\mu_{1}\nu_{1}...\mu_{n}\nu_{n}}R_{\rho_{1}\sigma_{1}}^{\mu_{1}\nu_{1}}...R_{\rho_{n}\sigma_{n}}^{\mu_{n}\nu_{n}}
+\nonumber\\&&
\sum_{n=1}^{p}m_{g}^{2n}\delta^{\rho_{1}\sigma_{1}...\rho_{n}\sigma_{n}}_{\mu_{1}\nu_{1}...\mu_{n}\nu_{n}}\kappa^{\mu_{1}}_{\rho_{1}}\kappa^{\nu_{1}}_{\sigma_{1}}....\kappa^{\mu_{n}}_{\rho_{n}}\kappa^{\nu_{n}}_{\sigma_{n}}+...\Big)]
\label{a27}
\end{eqnarray}

In above equation, second term corresponds to Lovelock gravity
\cite{w22,w23} and third term is related to nonlinear massive
gravity. Also, there are other terms that may produce another
types of gravity theories. These results show that by increasing
the number of dimensions of branes, more wormholes are created and
the interaction between branes increases which causes to emergence
of different gravity terms in relevant action of branes.

\section{Unifying nonlinear gravity with nonlinear electrodynamics by compacting Mp-branes}\label{o2}
Until now, we have shown that Mp-branes are constructed from p
M1-branes. Each of these branes are connected by M1-branes of
other Mp-brane by a wormhole. By adding the number of dimensions,
the number of nonlinear terms in relevant action of branes
increases and some theories like lovelock and nonlinear gravity
are produced. At this stage, we can show that graviton has a
direct relation with photon. In fact, by compacting Mp-branes, it
may transit to Dp-brane and some gravitons convert to photons
\cite{www12,ww13,ww14}. To show this, we use of the method in
\cite{w7,w20} and define $<X^{10}>=\frac{R}{l_{p}^{3/2}}$ where
$l_{p}$ is the Planck length. We can write:

\begin{eqnarray}
&& [X^{a},X^{b},X^{c}]=F^{abc} \quad [X^{a},X^{b}]=F^{ab} \nonumber \\
&& \nonumber \\
&&F_{abc}=\partial_{a} A_{bc}-\partial_{b} A_{ca}+\partial_{c}
A_{ab} \quad F_{ab}=\partial_{a} A_{b}-\partial_{b} A_{a}\nonumber \\
&& \nonumber \\
&& \nonumber \\
&&\Sigma_{a,b,c=0}^{10} \langle
F^{abc},F_{abc}\rangle=\Sigma_{a,b,c=0}^{10}
\langle[X^{a},X^{b},X^{c}],[X_{a},X_{b},X_{c}]\rangle=
 \nonumber \\
&& - \Sigma_{a,b,c,a'b'c'=0}^{10}\varepsilon_{abcD}\varepsilon_{a'b'c'G}^{D}X^{a}X^{b}X^{c}X_{a'}X_{b'}X_{c'} = \nonumber \\
&& - 6\Sigma_{a,b,a',b'=0}^{9}\varepsilon_{ab10D}\varepsilon_{a'b'10}^{D}X^{a}X^{b}X^{10}X_{a'}X_{b'}X_{10} - \nonumber \\
&& 6\Sigma_{a,b,c,a',b',c'=0,\neq 10}^{9}\varepsilon_{abcD}\varepsilon_{a'b'c'}^{D}X^{a}X^{b}X^{c}X_{a'}X_{b'}X_{c'} = \nonumber \\
&& - 6(\frac{R^{2}}{l_{p}^{3}})\Sigma_{a,b,a',b'=0}^{9}\varepsilon_{ab10D}\varepsilon_{a'b'10}^{D}X^{a}X^{b}X_{a'}X_{b'} - \nonumber \\
&& 6\Sigma_{a,b,c,a',b',c'=0,\neq10}^{9}\varepsilon_{abcD}\varepsilon_{a'b'c'}^{D}X^{a}X^{b}X^{c}X_{a'}X_{b'}X_{c'} = \nonumber \\
&& - 6(\frac{R^{2}}{l_{p}^{3}})\Sigma_{a,b=0}^{9}[X^{a},X^{b}]^{2} - \nonumber \\
&& 6\Sigma_{a,b,c,a',b',c'=0,\neq10}^{9}\varepsilon_{abcD}\varepsilon_{a'b'c'}^{D}X^{a}X^{b}X^{c}X_{a'}X_{b'}X_{c'}= \nonumber \\
&& \nonumber \\
&& - 6(\frac{R^{2}}{l_{p}^{3}})\Sigma_{a,b=0}^{9}F^{ab}F_{ab} +
E_{Extra} \label{a28}
\end{eqnarray}

This equation shows that self energy of gravitons in eleven
dimensional space-time is equal to self energy of photons in ten
dimensional space-time in additional to some energy which is
produced as due to compactification of Mp-branes. Using equation
(\ref{a28}), we can replace all terms in gravity theories by terms
in electrodynamics.

\begin{eqnarray}
&&\Sigma_{\rho,\sigma,\mu,\nu=0}^{10}R^{\rho}_{\sigma\mu\nu}=\Sigma_{\rho,\sigma,\mu,\nu,\lambda=0}^{10}\langle
F^{\rho}\smallskip_{\sigma\lambda},F^{\lambda}\smallskip_{\mu\nu}\rangle=\Sigma_{\rho,\sigma,\mu,\nu,\lambda=0}^{10}\langle[X^{\rho},X_{\sigma},X_{\lambda}],[X^{\lambda},X_{\mu},X_{\nu}]\rangle\Rightarrow
\nonumber\\&&
6(\frac{R^{2}}{l_{p}^{3}})\Sigma_{\rho,\sigma,\mu,\nu=0}^{9}[X^{\rho},X_{\sigma}][X_{\mu},X_{\nu}]=6(\frac{R^{2}}{l_{p}^{3}})
\Sigma_{\rho,\sigma,\mu,\nu=0}^{9}F^{\rho}\smallskip_{\sigma}F_{\mu,\nu}\nonumber\\&&
\Sigma_{\rho,\sigma,\mu,\nu=0}^{10}R=\Sigma_{\rho,\sigma,\lambda=0}^{10}\langle
F^{\rho\sigma}\smallskip_{\lambda},F^{\lambda}\smallskip_{\rho\sigma}\rangle=
\Sigma_{\rho,\sigma,\lambda=0}^{10}\langle[X^{\rho},X^{\sigma},X_{\lambda}],[X^{\lambda},X_{\rho},X_{\sigma}]\rangle\Rightarrow
\nonumber\\&&
6(\frac{R^{2}}{l_{p}^{3}})\Sigma_{\rho,\sigma=0}^{9}[X^{\rho},X^{\sigma}][X_{\rho},X_{\sigma}]=6(\frac{R^{2}}{l_{p}^{3}})
\Sigma_{\rho,\sigma=0}^{9}F^{\rho\sigma}F_{\rho\sigma}\label{a29}
\end{eqnarray}

These relations help us to replace Lovelock theory with nonlinear
electrodynamics:

\begin{eqnarray}
&&
\sum_{n=1}^{p}m_{g}^{2n}\delta^{\rho_{1}\sigma_{1}...\rho_{n}\sigma_{n}}_{\mu_{1}\nu_{1}...\mu_{n}\nu_{n}}
R_{\rho_{1}\sigma_{1}}^{\mu_{1}\nu_{1}}...R_{\rho_{n}\sigma_{n}}^{\mu_{n}\nu_{n}}\Rightarrow
\nonumber\\&&\sum_{n=1}^{p}m_{g}^{2n}(\frac{R^{2n}}{l_{p}^{3n}})
\delta^{\rho_{1}\sigma_{1}...\rho_{n}\sigma_{n}}_{\mu_{1}\nu_{1}...\mu_{n}\nu_{n}}
F^{\mu_{1}\nu_{1}}...F^{\mu_{n}\nu_{n}}F_{\rho_{1}\sigma_{1}}..F_{\rho_{n}\sigma_{n}}\label{a30}
\end{eqnarray}

This equation shows that by compacting Mp-brane, nonlinear
theories like Lovelock gravity converts to nonlinear
electrodynamics. In fact, the origins of Lovelock terms and
nonlinear electromagneic field \cite{ww12,ww13,ww14} are the same.
We can consider other effects of compactifications of Mp-brane by
extending the relations in (\ref{a29}) and writing following
relations:

\begin{eqnarray}
&&\Sigma_{a,i,j=0}^{10}\langle[X^{a},X^{i},X^{j}],[X_{a},X_{i},X_{j}]\rangle\Rightarrow
\nonumber\\&&
6(\frac{R^{2}}{l_{p}^{3}})\Sigma_{i,j=0}^{9}[X^{i},X^{j}][X_{i},X_{j}]\nonumber\\
&&\Sigma_{a,b,i=0}^{10}\langle[X^{a},X^{i},X^{b}],[X_{a},X_{i},X_{b}]\rangle\Rightarrow
\nonumber\\&&
6(\frac{R^{2}}{l_{p}^{3}})\Sigma_{i,b=0}^{9}[X^{i},X^{b}][X_{i},X_{b}]\nonumber\\
&&\Sigma_{a,b,i=0}^{10}\langle[X^{a},X^{b},X^{i}],[X_{a},X_{b},X_{i}]\rangle\Rightarrow
\nonumber\\&&
6(\frac{R^{2}}{l_{p}^{3}})\Sigma_{i,b=0}^{9}[X^{b},X^{i}][X_{b},X_{i}]\label{a31}
\end{eqnarray}

On the other hand, we can write  following mappings for
noncommutative brackets in ten dimensional space-time \cite{w7}:

\begin{eqnarray}
&&\langle[X^{a},X^{i}],[X^{a},X^{i}]\rangle=
\frac{1}{2}(\partial_{a}X^{i})(\partial_{a}X^{i}) \nonumber \\
&&\nonumber \\
&&\langle[X^{a},X^{b}],[X^{a},X^{b}]\rangle=\langle
F^{ab},F^{ab}\rangle \nonumber \\
&&\nonumber \\
&& i,j=p+1,..,10\quad a,b=0,1,...9\quad m,n=0,..,10~~ \label{a32}
\end{eqnarray}

Using the relations in equations
(\ref{a28}),(\ref{a29}),(\ref{a30}),(\ref{a31}) and (\ref{a32}) in action
(\ref{a4}), we can obtain the relevant action for D3-brane:

\begin{eqnarray}
&&S_{D3} = -T_{D3} \int d^{4}\sigma
\Big(\frac{1}{2}\sum_{a=0,..3}\sum_{i=4..9}(\partial_{a}X^{i})(\partial_{a}X^{i})
+6(\frac{R^{2}}{l_{p}^{3}})\Sigma_{i,j=4}^{9}[X^{i},X^{j}][X_{i},X_{j}]-\nonumber \\
&&\sum_{n=1}^{3}m_{g}^{2n}(\frac{R^{2n}}{l_{p}^{3n}})
\delta^{\rho_{1}\sigma_{1}...\rho_{n}\sigma_{n}}_{\mu_{1}\nu_{1}...\mu_{n}\nu_{n}}
F^{\mu_{1}\nu_{1}}...F^{\mu_{n}\nu_{n}}F_{\rho_{1}\sigma_{1}}..F_{\rho_{n}\sigma_{n}}\Big)^{1/2}\label{a33}
\end{eqnarray}

This equation is in agreement with relevant action for D3-branes
\cite{w7,w12,w13,w14,w15,w16} if we replace the nonlinear term of
electrodynamics  by linear one. As can be seen from the above
equation, the nonlinear gravity in M-theory can be replaced by
nonlinear electrodynamics in string theory. This is because that
the origin of all matters are strings. When strings are attached
to Mp-branes, they have the role of gravitons. By compacting
Mp-branes and reducing them to Dp-branes, they may play the role
of nonlinear photons and in conditions that they are deliberated
from branes, they can be regarded as scalars.

\section{Summary and Discussion} \label{sum}
In this paper, we have shown that  Mp-branes can be built by
joining and growing  of p M1-branes. These M1-branes may be
connected to M1-branes of other Mp-branes by a wormhole. Thus,
each Mp-brane may have p wormholes which originate from nonlinear
gravity on it. In fact, for an M1-brane, we only have linear
gravity; however by gluing M1-branes and emergence of Mp-branes,
the  nonlinear terms in relevant action of them are produced and
some theories like lovelock and nonlinear gravity are born. By
compacting Mp-brane, nonlinear theories converts to nonlinear
electrodynamics. This is because that the origin of graviton and
photon are strings. When strings glue to Mp-branes, they have the
role of gravitons. By compacting Mp-branes and reducing them to
Dp-branes, they may play the role of nonlinear photons and in
conditions that they are deliberated from branes, they can be
regarded as  scalars.

\section*{Acknowledgments}
\noindent  A. Sepehri would like to thank of the Research
Institute for Astronomy and Astrophysics of Maragha, Iran for
financial support during investigation in this work.

 \end{document}